\documentclass[useAMS,usenatbib]{mn2e}
\usepackage{graphicx}  
\usepackage{float}     
\usepackage{amsmath}

\title[The Evolution of Interacting Spiral Galaxy NGC\,5194]
      {The Evolution of Interacting Spiral Galaxy NGC\,5194}

\author[X. Y. Kang et al. ]
{Xiaoyu~Kang$^{1,2,4}$\thanks{E-mail: kxyysl@ynao.ac.cn},
 Ruixiang~Chang$^{3}$\thanks{E-mail: crx@shao.ac.cn},
 Fenghui~Zhang$^{1,2}$\thanks{E-mail: zhangfh@ynao.ac.cn},
 Liantao~Cheng$^{1,2,4}$,
 Lang~Wang$^{1,2,4}$\\
$^1$National Astronomical Observatories, Yunnan Observatory, Chinese Academy of Sciences, Kunming, 650011, China\\
$^2$Key Laboratory for the Structure and Evolution of Celestial Objects, Yunnan Astronomical Observatory, Chinese Academy of Sciences,\\
 Kunming, 650011, China\\
$^3$Key Laboratory for Research in Galaxies and Cosmology, Shanghai Astronomical Observatory, Chinese Academy of Sciences, 80 Nandan \\
Road, Shanghai, 200030, China \\
$^4$University of Chinese Academy of Sciences, Beijing 100049, China\\
   }
\begin{document}


\pagerange{\pageref{firstpage}--\pageref{lastpage}}
\pubyear{2014}

\maketitle

\label{firstpage}

\begin{abstract}


NGC\,5194 (M51a) is a grand-design spiral galaxy and
undergoing interactions with its companion. Here
we focus on investigating main properties of its
star-formation history (SFH) by constructing a simple evolution
model, which assumes that the disc builds up gradually
by cold gas infall and the gas
infall rate can be parameterizedly described
by a Gaussian form. By comparing model predictions with
the observed data, we discuss the probable range
for free parameter in the model and then know more
about the main properties of the evolution and SFH of M51a.
We find that the model predictions are very sensitive
to the free parameter and the model adopting a
constant infall-peak time $t_{\rm p}\,=\,7.0{\rm Gyr}$
can reproduce most of the observed constraints of M51a.
Although our model does not assume the gas infall time-scale
of the inner disc is shorter than that of the outer disc, our model
predictions still show that the disc of M51a
forms inside-out. We find that the mean stellar age
of M51a is younger than that of the Milky Way, but older
than that of the gas-rich disc galaxy UGC\,8802. In
this paper, we also introduce a 'toy' model to allow
an additional cold gas infall occurred recently
to imitate the influence of the interaction
between M51a and its companion. Our results show that
the current molecular gas surface density, the SFR and the UV-band
surface brightness are important quantities to trace
the effects of recent interaction on galactic SF process.

\end{abstract}


\section{Introduction}

M51a is one of the nearest galaxies with grand-design
spiral arms. The outstanding point of this galaxy is that it is
known to be interacting with its companion galaxy NGC\,5195 (the
interacting system NGC\,5194 + NGC\,5195 is often called as M51).
Coupled with its large angular size, nearly face-on inclination, high
surface brightness and kinematic complexity, many studies about M51a
have been carried out using various wavelengths and methods.

With respect to dynamic study, M51a is a very good target to explore
the nature and origin of spiral structure. Two opposite
scenarios dominate the discussion in the literature. One theory considers that
the spiral arms are a long lasting pattern that slowly evolves and
rotates with a single angular speed \citep{lin&shu}, while the other
assumes the arms to be transient disturbances generated, e.g., by the
tidal interaction with a companion \citep{Toomre2}. In the
seminal kinematic study of M51a, \citet{tully74} identified the spiral
pattern in the outer disc as transient feature, while the inner arms
are thought to be in a steady state. Indeed, \citet{Colombo14}
showed the kinematic evidence of an $m\,=\,3$ wave in the inner disc of
M51a, which may suggest the density-wave nature of the main
spiral structure of M51a. On the other hand, many kinematic and
hydrodynamical models showed evidence that, instead of quasi-steady
density wave, the internal structure of M51a drives from the
complicated and dynamical interaction with its companion
\citep{dobbs10, S&L00, Theis03}.

In addition to its dynamics, M51a is also a test bed to understand
the effects of galaxy interacting on its star formation (SF) process by
investigating its stellar populations. Early observations have
already shown that the blue and yellow-red stellar populations
within M51 are spatially separate structures \citep{Zwicky55}. Later
investigations of star clusters in M51a revealed that the star
cluster formation rate increased significantly during the period
of 100-250\,Myr ago, which is consistent with the epoch of the
dynamic encounters of these two galaxies \citep{lee05, Hwang08, Hwang10}.
\citet{Kaleida10} also found that there was an enhancement
in the number of young stellar associations in the northern
arm closest to the companion, most likely triggered
by the interaction of M51a and NGC 5195. Through the spectral
energy distribution modeling of the multi-bands images of
the M51 system, \citet{cooper12} revealed a burst of star
formation occurred in both galaxies roughly 340-500\,Myr ago,
which is in agreement with the results of the colour-magnitude
diagrams of individual stars \citep{tikhonov09}. Moreover,
\citet{lee11} found evidence that the tidal interaction
between M51a and its companion appeared to enhance the SF
process at the tidal bridge connecting the two galaxies.

On the other hand, although the parameterized models have already been
applied to several individual galaxies and proven to be fruitful
tools to explore galactic formation and evolution \citep{Tinsley80,
chang99,chang12,bp00,chiappini01,yin09,kang12}, there is still
lack of similar investigation on M51a partially due to the complexity
caused by the on-going interacting between M51a and its companion.
In this paper, we construct a simple evolution model for M51a to
build a bridge between its SFH and its observed
properties, especially the radial distributions of cold gas surface density,
metallicity, and the radial profiles of surface brightness in multi-bands.
We present a parameterized description of the gas cooling process of
the disc of M51a and adopt the local SF law in the model
to calculate its star formation rate (SFR) using the cold gas surface density.
Then, our model can predict the SFH, the chemical
and colour evolution of M51a, with the help of stellar population
synthesis (SPS) method. By comparing model predictions with
the observed data, we can discuss the probable ranges for free
parameters in the model and then know more about the main properties
of the evolution and SFH of M51a.

The paper is structured as follows. Section 2 describes the observed
features of M51a disc, including the surface brightness, the cold gas
content, the SFR and the metallicity. In Section 3 we
present the main assumptions and ingredients of our
model in details. The comparisons between model predictions
and observations and our main results are shown in
Section 4. Our main conclusions are presented in Section 5.

\section{Observations}
\label{sect:Obs}

M51a is one of the closest
\citep[the distance: $\rm D\sim8.0\,Mpc$;][]{walter08}, face-on
\citep[inclination: $i\sim20^{\circ}$;][]{tully74} SA(s)bc
spiral galaxy and many of its basic quantities have already
been measured \citep{hu13}. A summary of the main properties of
M51a is shown in Table \ref{Tab:obs}, most of which
are adopted to constrain our model.

In this Section, we summarize the current available
observations of M51a, especially the radial
distributions along the disc, including the surface
densities of gas mass and SFR, surface brightness and
metallicity.

\begin{table}
\caption{The main properties of M51a.}
\label{Tab:obs}
\begin{center}
\begin{tabular}{lll}
\hline
\hline
Hubble type          &  SA(s)bc                                 & 1\\
Distance             & 8.0~Mpc                                 & 2\\
Inclination          & $22^{\circ}$                      & 3\\
$M_{\rm K}$           & -24.19\,mag                      & 4 \\
Stellar mass         & $\sim3.6\times10^{10}\,{\rm M_{\odot}}$         & 5\\
H{\sc i} mass        & $\sim(2.8-3.9)\times10^{9}\,{\rm M_{\odot}}$     & 5, 6 \\
H$_2$ mass           & $\sim(2.52-7.1)\times10^9\,{\rm M_{\odot}}$        & 5, 6\\
$f_{\rm gas}=M_{\rm gas}/(M_{\rm gas}+M_{\rm star})$         & $\sim(0.13-0.23)$   &  \\
SFR$_{\rm FUV+24\mu m}$      & $3.125~{\rm M_{\odot }}$\,yr$^{-1}$    &5\\
SFR$_{\rm H\alpha }$ & $5.4~{\rm M_{\odot }}$\,yr$^{-1}$              & 7\\
SFR$_{\rm RC-20cm}$  & $2.56~{\rm M_{\odot }}$\,yr$^{-1}$              & 8\\
\hline
\end{tabular}\\
\end{center}
Refs: (1) NED; (2) \citet{walter08}; (3) \citet{tully74}; (4) \citet{jarrett03};
(5) \citet{leroy08}; (6) \citet {miyamoto14}; (7) \citet{kennicutt03};
(8) \citet{schuster07}
\end{table}

\subsection{Radial profiles of surface brightness}
\label{sect:sb}

The radial surface brightness profiles in multi-band
of a spiral galaxy contain fossil information about
the SFH along the galactic disc and provide strong
constraints on the models of galactic formation
and evolution.

The catalog data of the radial profiles of surface brightness
for M51a in the ultraviolet bands (UV; \textit{Galaxy Evolution
Explorer, GALEX} for short), optical bands (Sloan Digital Sky Survey,
SDSS for short), and near-infrared bands (NIR; Two-Micron All-Sky Survey,
2MASS for short) have been published by \citet[][MM09I hereafter]{MM091},
 but all these values were only corrected for foreground Galactic
extinction and still in lack of internal extinction correction.
We adopt the method presented by \citet{linl13} to
estimate the internal extinction profile in the FUV band $A_{\rm FUV}$.
Then, following the prescriptions of \citet{boselli03} and
\citet{cortese08}, we calculate the internal extinction
in other wavelengths by assuming a given extinction law
and a geometry for the distribution of stars and dust.
The radial surface brightness profiles
in FUV-, NUV-, $u$-, $g$-, $r$-, $i$-, $z$-, $J$-, $H$- and
$K$-band are plotted in the left-hand side of
Fig. \ref{Fig:results1} and Fig. \ref{Fig:results2},
where the open circles show the observed profiles only
corrected for the Galactic extinction and the filled
ones also include a correction for the radial variation of internal
extinction.

The surface brightness in the $K$-band has another application.
Since the $K$-band luminosity is most sensitive to the old stellar
population and less effected by dust attenuation than other bands,
we adopt the $K$-band mass-to-light ratio to be
$\Upsilon_{\star}^{K}\,=\,0.5\,{\rm M_{\odot}}/{\rm L}_{\odot,K}$
\citep{bell03,leroy08}, and then utilize the observed $K$-band surface
brightness profile \citep{MM091} to estimate the present-day
stellar mass surface density profile $\Sigma_*(r,t_{\rm g})$,
where $t_{\rm g}$ is the cosmic age and we set
$t_{\rm g}\,=\,13.5\rm\,Gyr$ according to our adopted cosmology
(see Section \ref{sect:analysis}). We will use $\Sigma_*(r,t_{\rm g})$
derived here as our primary input parameter to constrain
the total mass infalling to the disc.


\subsection{Radial profiles of cold gas and SFR}

During the past years, a number of data sets relating to the atomic
and molecular gas surface density in M51a are becoming available.

In spirals, the CO line emissions are usually used to estimate
the molecular gas mass surface density $\Sigma_{\rm H_2}$.
Imaging of molecular clouds has recently been carried out in M51a
with the Berkeley-Illinois-Maryland Association (BIMA)
interferometer using the CO $J=1-0$ line transition \citep{helfer03}.
Observations of the CO $J=2-1$ emission in M51a were conducted with
the Institute de RadioAstronomie Millim\'{e}trique (IRAM) 30m
telescope using the 18 element focal plane heterodyne receiver
array HERA \citep{schuster04}. Observations of CO $J=1-0$
emission for M51a were also carried out using the 45-m
telescope of the Nobeyama Radio Observatory (NRO) with
the $5\times5-$beam SIS heterodyne receiver array (BEARS).
\citet{schuster07} derived $\Sigma_{\rm H_2}$ with the help
of the CO $J\,=\,2-1$ line from HERA focal plane array, while
others used the CO $J\,=\,1-0$ line maps from BIMA and BEARS
to obtain $\Sigma_{\rm H_2}$ \citep{kennicutt07,leroy08,miyamoto14}.

To derive the surface density of neutral atomic hydrogen of M51a,
\citet{kennicutt07}, \citet{leroy08} and
\citet{miyamoto14} used Very Large Array (VLA)
maps of the 21\,cm line obtained as part of The H{\sc i}
Nearby Galaxy Survey (THINGS) \citep{walter08}.
\citet{schuster07} estimated the corresponding surface density
from the large-scale distribution of the 21c\,m line of atomic
hydrogen in M51a analyzed by \citet{rots90} using the VLA.

The sum of the surface density of atomic and molecular gas
is often called as the total gas surface density, i.e.,
$\Sigma_{\rm{gas}}\,=\,1.36(\Sigma_{\rm{H_{2}}}+\Sigma_{\rm HI})$,
and the factor 1.36 is to include the contribution of helium.
The observed radial distributions of molecular, atomic and total
gas mass surface density are plotted in the right-hand
side of Fig. \ref{Fig:results1} and Fig. \ref{Fig:results2},
where the plotted data are taken from \citet{schuster07} (filled circles),
\citet{kennicutt07} (filled triangles), \citet{leroy08}
(asterisks) and \citet{miyamoto14} (filled diamonds).

With respect to the radial distribution of the SFR, several
groups have already measured the SFR surface density,
$\Sigma_{\rm SFR}$, in the disc of M51a using different tracers.
The radio-continuum at 20\,cm wavelength is used to estimate
$\Sigma_{\rm SFR}$ by \citet{schuster07}.
\citet{kennicutt07} used Pa$\alpha$ and a combination
of $24\,\mu\rm m$ and H$\alpha$ emission to obtain $\Sigma_{\rm SFR}$.
\citet{leroy08} combined FUV and  $24\,\mu\rm m$ maps to
estimate $\Sigma_{\rm SFR}$.
\citet{heesen14} obtained the radial distribution of the
$\Sigma_{\rm SFR}$ from $\lambda22\,\rm cm$ radio-continuum
emission.

The right fourth panel of Fig. \ref{Fig:results1} and Fig. \ref{Fig:results2}
plots the radial profiles of SFR surface density, $\Sigma_{\rm SFR}$.
The data taken from \citet{schuster07} and \citet{heesen14} are
shown as filled circles ($\lambda20\,\rm cm$ RC emission)
and the filled pentagons ($\lambda22\,\rm cm$
RC emission), respectively. The data obtained from \citet{leroy08}
are represented by the the filled asterisks
(24$\,\mu \rm m\,+\,FUV$), while that from \citet{kennicutt07} are
denoted by filled triangles (24$\,\mu\rm m\,+\,H\alpha$).

\subsection {Radial profiles of metallicity}

The observed metallicity gradient is another important constraint on
the models of the galactic evolution. Except for hydrogen and helium,
oxygen is the most abundant element in the Universe \citep{korotin14},
and oxygen abundance is easily measured in H{\sc ii} regions
because of its bright emission line.
In practice, the oxygen abundance is sometimes used
to represent the metallicity of the galaxy.

The observed oxygen abundance gradients of H{\sc ii} regions in M51a
disc have been estimated by several authors. Based on analysis of
a sample of 10 H{\sc ii} regions with high-quality electron temperatures
measurements, \citet{bresolin04} derived an oxygen radial gradient of
$-0.021\pm0.011\rm\,dex\,kpc^{-1}$. \citet{moustakas10} adopted
two calibration methods to convert line strength to oxygen
abundance and obtained two values of radial gradient. One is
$-0.038\pm0.004\rm\,dex\,kpc^{-1}$ by using the
theoretical calibration method of \citet[][KK04 hereafter]{KK04} and
the other is $-0.024\pm0.004\rm\,dex\,kpc^{-1}$ by adopting the empirical
calibration method of \citet[][PT05 hereafter]{PT05}.
\citet{P14} obtained an oxygen radial gradient of
$-0.0223\pm0.0037\rm\,dex\,kpc^{-1}$.

The observed oxygen abundance profile of the M51a disc is shown
in the right bottom panel of Fig. \ref{Fig:results1} and Fig.
\ref{Fig:results2}. The cycles represent the observed data
obtained from \citet{moustakas10}, where the open and filled
circles are derived by the PT05 and KK04 calibration method,
respectively. In the same panel, the data from \citet{P14} are
shown as filled squares.

However, the true situation of the oxygen abundance gradient
in the M51a disc is still not fixed. The main uncertainty
comes from the calibration methods used to derive chemical
abundance. In addition, the data might suffer
from lack of large enough samples. In any case, a larger
and homogeneous sample of H{\sc ii} regions which spread
the whole M51a disc is needed in order to have
conclusive results about the real oxygen abundance gradients.


\section{The Model}
\label{sect:analysis}

Similar to previous models of the Milky Way \citep{chang99,
chiappini01, matteucci09} and other nearby disc galaxies
\citep{bp00, m&d05, dalcanton07,yin09, MM11, kang12, chang12, r&v13},
we assume that the M51a disc is progressively built up by the infall
of primordial gas ($X\,=\,0.7571, Y_{\rm p}\,=\,0.2429, Z\,=\,0$) from
its dark matter halo. The disc is basically assumed to be
sheet-like and composed of a series of independently evolved
rings with width 500 pc, in the sense that no radial mass flows are
considered.

Throughout this paper we adopt the standard cold dark matter
cosmology with $H_{0}\,=\,70\,\rm{km\,{s}^{-1}\,{Mpc}^{-1}}$,
$\Omega_{\rm M}\,=\,0.3$, and $\Omega_\Lambda\,=\,0.7$.
Correspondingly, the cosmic time at $z\,=\,0$ is
$13.5\,{\rm Gyr}$. The details and essentials of our
model are described as follows.

\subsection{Gas infall rate}

We assume that there is a $1\,\rm{Gyr}$ time delay for the disc
formation, which corresponds to $z\,\sim\,6$ under the standard
cosmology. After that, the disc originates and grows
by continuous primordial gas infall from the dark matter halo.
We adopt a Gaussian formula of the gas infall rate from
\citet{chang99, chang12}.

\begin{figure}
  \centering
  \includegraphics[angle=0,width=0.475\textwidth]{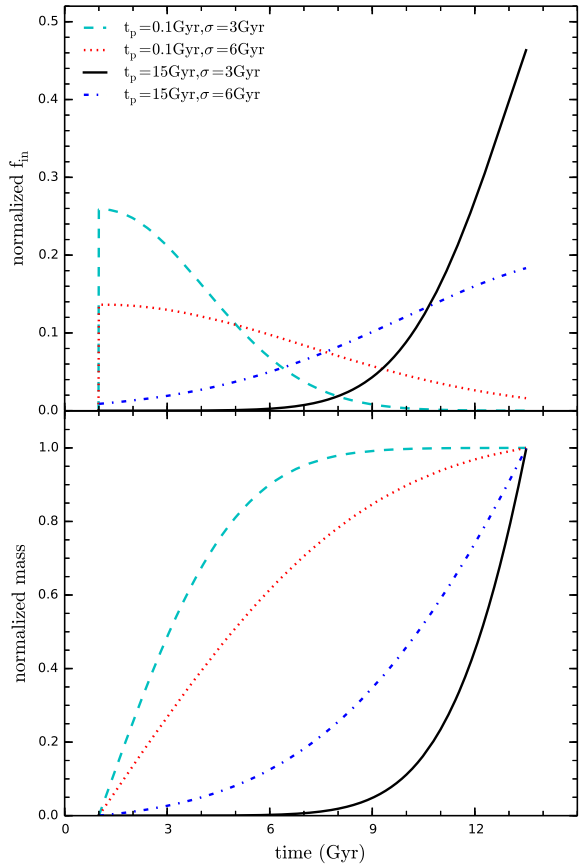}
    \caption{The normalized gas infall rate (Upper panel) and the total
    infall gas surface density (Lower panel).
    Different line types correspond to various parameter groups:
    dashed lines $(t_{\rm p}, \sigma)=(0.1\,{\rm Gyr}, 3.0\,{\rm Gyr})$,
    dotted lines $(t_{\rm p}, \sigma)=(0.1\,{\rm Gyr}, 6.0\,{\rm Gyr})$,
    dot-dashed lines $(t_{\rm p}, \sigma)=(15\,{\rm Gyr}, 6.0\,{\rm Gyr})$,
    solid lines $(t_{\rm p}, \sigma)=(15\,{\rm Gyr}, 3.0\,{\rm Gyr})$.
    }
  \label{Fig:sig}
\end{figure}

For the given radius $r$, the infall rate
$f_{\rm{in}}(r,t)$ (in units of $\rm{M_{\odot}}\,{pc}^{-2}\,{Gyr}^{-1}$)
is assumed to be:
\begin{equation}
f_{\rm{in}}(r,t)=\frac{A(r)}{\sqrt{2\pi}\sigma}e^{-(t-t_{\rm p})^{2}/2\sigma^{2}},
\label{eq:infall rate}
\end{equation}
where $t_{\rm p}$ is the infall-peak time and $\sigma$ is the full
width at the half-maximum of the peak. The $A(r)$ is a set of
normalized quantities constrained by the stellar mass surface density
at the present time $\Sigma_*(r,t_{\rm g})$. To explore how
$t_{\rm p}$ and $\sigma$ regulate the shape of the gas
infall rate, Fig. \ref{Fig:sig} plots the normalized gas infall
rate (upper panel) and the corresponding total infall
gas surface density (lower panel), which is defined as
how many cold gas have already been cooled to the given
ring at time $t$ and normalized to 1 at the present time.
Different line types correspond to various parameter
groups $(t_{\rm p}, \sigma)$. We select $t_{\rm p}=0.1\rm\,Gyr$
and $t_{\rm p}=15\rm\, Gyr$ to represent the extreme case
which corresponds to a time-decreasing
infall rate and a time-increasing one, respectively.
It can be seen from Fig. \ref{Fig:sig} that
the degeneracy of $t_{\rm p}$ and $\sigma$ does exit in that the
combination of small $t_{\rm p}$ and large $\sigma$ results in a
similar shape of mass accretion curve comparing that of large
$t_{\rm p}$ and small $\sigma$. Since our main aim is to investigate
the main trend of SFH for M51a but not to find the exact values of
free parameters, for the purpose of simplicity, we choose
to reduce the number of free parameters in our model
by treating $t_{\rm p}$ as the free parameter and fixing
$\sigma\,=\,3\,\rm Gyr$ in the following
section of this paper \citep{chang10, chang12}.

We should point out that, in previous works by
\citet{hou00}, \citet{chiappini01} and \citet{yin09},
an exponential form of gas infall rate
with one free parameter is widely adopted in models of galactic
chemical evolution. But, in these models,
the gas infall rate always decreases with time and the most extreme
case corresponds to a constant one. Therefore, we adopt a Gaussian
formula of infall rate in this paper to include more
possibilities, especially the time-increasing gas infall rate.

\begin{figure}
  \centering
  \includegraphics[angle=0,width=0.475\textwidth]{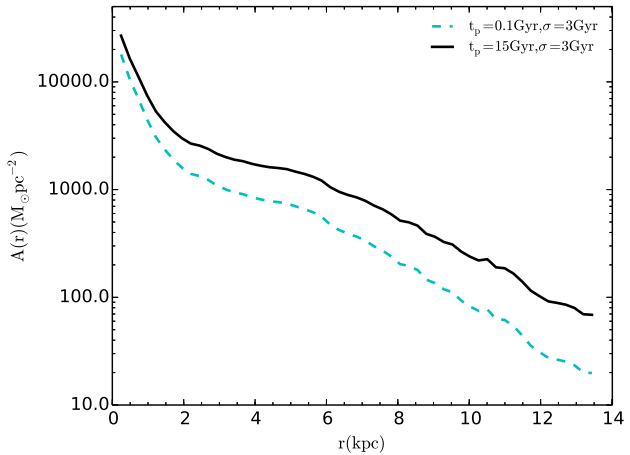}
    \caption{The normalized quantities $A(r)$. The dashed and solid lines
    are corresponding to two limiting cases of $t_{\rm p}=0.1\,\rm Gyr$
    and $t_{\rm p}=15\,\rm Gyr$, respectively.
    }
  \label{Fig:A_r}
\end{figure}

The normalized quantities $A(r)$ in Equ. (\ref{eq:infall rate})
control how many cold gas finally cool to the disc
as a function of radius. In practice, we iteratively estimate $A(r)$
by requiring the model resulted stellar mass surface
density at the present time is equal to its observed value
\citep{chang10,chang12,kang12}. In our calculation, after
fixing the values of $t_{\rm p}$ and $\sigma$, we give an
initial value of $A(r)$ and numerically
calculate the gas evolution and the increase of stellar mass.
By comparing the resulted $\Sigma_*(r,t_{\rm g})$ with its
observed value, we adjust the value of A(r) and repeat
the calculation untill the resulted $\Sigma_*(r,t_{\rm g})$
is comparable to its observed value. Fig. \ref{Fig:A_r} plots
the resulted radial profile of A(r), where the dashed and solid lines
are corresponding to two limiting cases of $t_{\rm p}=0.1\,\rm Gyr$
and $t_{\rm p}=15\,\rm Gyr$, respectively. It shows that the
A(r) does vary with $t_{\rm p}$. But we do not treat A(r) as
free parameter in our model since it is roughly fixed for given
$t_{\rm p}$.

Regarding the stellar mass profile at the present time, in
our previous works such as \citet{chang12} and
\citet{kang12}, we assumed that the stellar disc is a pure
exponential disc and the value of $\Sigma_*(r,t_{\rm g})$ is constrained
by the quantities of total stellar mass and radial scale-length,
which are usually obtained from the observed radial
profiles of surface brightness plusing the exponential disc assumption.
In this paper, we regard $\Sigma_*(r,t_{\rm g})$ as a set of individual
quantities and can be directly estimated from the observed radial
profile of $K$-band surface brightness by adopting a constant
mass-to-light ratio (see Section \ref{sect:sb} for details).
Since this method includes much more spatial structures
than the previous ones, it permits us to more reasonably calculate
the multi-band radial profiles of surface brightness and compare
with the observations in details.

\subsection{Star formation law}

SF process plays an important role in regulating the
evolution of galaxies. Unfortunately, the SF process
is so complicated that its underlaying physical nature
is still not completely understood \citep{kennicutt07}.
Most models of the galactic evolution still rely on
the empirical SF law, which connects the local cold gas
surface density to the SFR surface density, to describe how
much cold gas converts into stellar mass in each time
step \citep{bp00,chiappini01,yin09,chang12,kang12}.

Based on the observed data of a sample of 97 nearby normal
and star-burst galaxies, \citet{k98} found a power-law correlation
between the galaxy-averaged SFR surface density and
the galaxy-averaged total gas surface density,
which was termed as the classical Schmidt-Kennicutt SF law
and widely adopted in the studies of galactic formation
and evolution. But a more fundamental way is to use spatially
resolved measurements of the SFR and gas surface
density to examine the correlations between the observables
on a point-by-point basis within galaxies. Indeed, the studies of
the spatially resolved SF law have been applied to several
nearby spiral galaxies \citep{kennicutt07,leroy08,bigiel08,
schruba11}. In the case of M51a, using the observed multiwavelength
data, \citet{schuster07} and \citet{kennicutt07} independently
investigated the spatially resolved SF law along the disc of M51a.
\citet{kennicutt07} found that the resolved SFR versus gas
surface density relation is well represented by a Schmidt power law,
which is similar in form and dispersion to the disc-average
Schmidt-Kennicutt law.

In our model, we adopt the SF law of M51a presented by
\citet{kennicutt07}:
\begin{equation}
\Psi(r,t)=0.048\Sigma_{\rm{gas}}(r,t)^{1.56},
\label{eq:sfr}
\end{equation}
where $\Psi(r,t)$ (in units of $\rm{M_{\odot}}\,{pc}^{-2}\,{Gyr}^{-1}$)
is the SFR surface density, and $\Sigma_{\rm gas}(r,t)$ (in units of
$\rm{M_{\odot}}\,{pc}^{-2}$) is the gas surface density, which is the
sum of the surface density of the atomic and molecular gas.
We assume that $\Sigma_{\rm{gas}}(r,t)\,=\,1.36
(\Sigma_{\rm{H_{2}}}(r,t)+\Sigma_{\rm HI}(r,t))$, where
$\Sigma_{\rm{H_{2}}}(r,t)$ and $\Sigma_{\rm HI}(r,t)$ are the
surface density of atomic and molecular hydrogen, and the factor 1.36 is
to reflect the contribution of helium.

Regarding the ratio of molecular-to-atomic gas surface density
$R_{\rm mol}(r,t)$, we adopt the formula presented by \citet{blitz06} and
\citet{leroy08},
\begin{equation}
R_{\rm{mol}}(r,t)\,=\,{\Sigma_{{{\rm{H}}_2}}}(r,t)/{\Sigma_{{\rm{H{_I}}}}(r,t)}\,=\,
{\left[ {P_{\rm h}\left( r,t\right)/{P_0}} \right]^{\alpha_P} },
\label{eq:BRh2}
\end{equation}
where $P_{\rm h}(r,t)$ is the mid-plane pressure of the interstellar medium
(ISM), $P_{0}$ and $\alpha_{\rm P}$ are constants derived from
the observations. We adopt
$P_{0}/k=1.92\times10^{4}\,\rm{cm}^{-3}\,\rm{K}$ and $\alpha_{\rm
P}=0.87$ derived by \citet{hitschfeld09} for M51a.

The mid-plane pressure of the ISM in disc galaxies can be expressed as
\citep{elmegreen89_1, leroy08}:
\begin{equation}
P_{\rm h}\left( r,t \right)\,=\,\frac{\pi }{2}\rm G{\Sigma _{{\rm{gas}}}}\left( r,t \right)\left[ {{\Sigma _{{\rm{gas}}}}\left( r,t \right)
+ \frac{c_{\rm gas}}{c_{\rm *}}{\Sigma _*}\left( r,t \right)} \right],
\label{eq:elmegreenpressure}
\end{equation}
where $G$ is the gravitational constant, and $c_{\rm{gas}}$ and
$c_{\rm{*}}$ are the (vertical) velocity dispersions of gas and
stars, respectively. Observations reveal that $c_{\rm{gas}}$ is a
constant along the disc and we adopt
$c_{\rm{gas}}\,=\,11\rm\,km\,s^{-1}$ \citep{Ostriker10}, but
$c_{\rm{*}}$ is estimated as
$c_{\rm *}\,=\,\sqrt{\pi Gz_{0}\Sigma _*(r)}$, where $z_{0}$ is the
scale-height of the disc and we adopt $z_{0}\,=\,1.0\,\rm{kpc}$
\citep{hitschfeld09}.

\subsection{Other ingredients and basic equations}

\begin{figure*}
   \centering
   \includegraphics[angle=0,height=18cm,width=11cm]{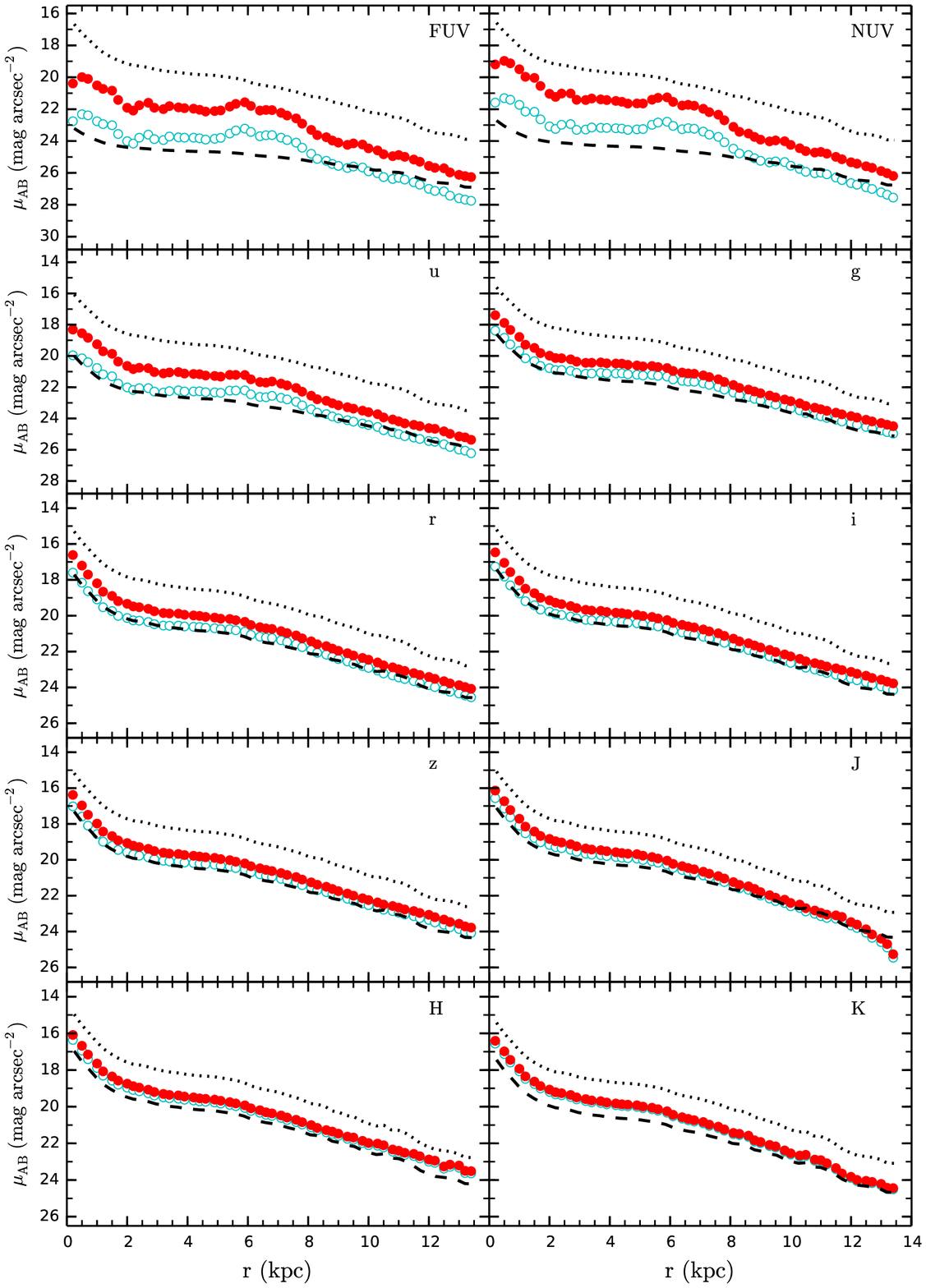}
   \includegraphics[angle=0,height=18cm,width=6cm]{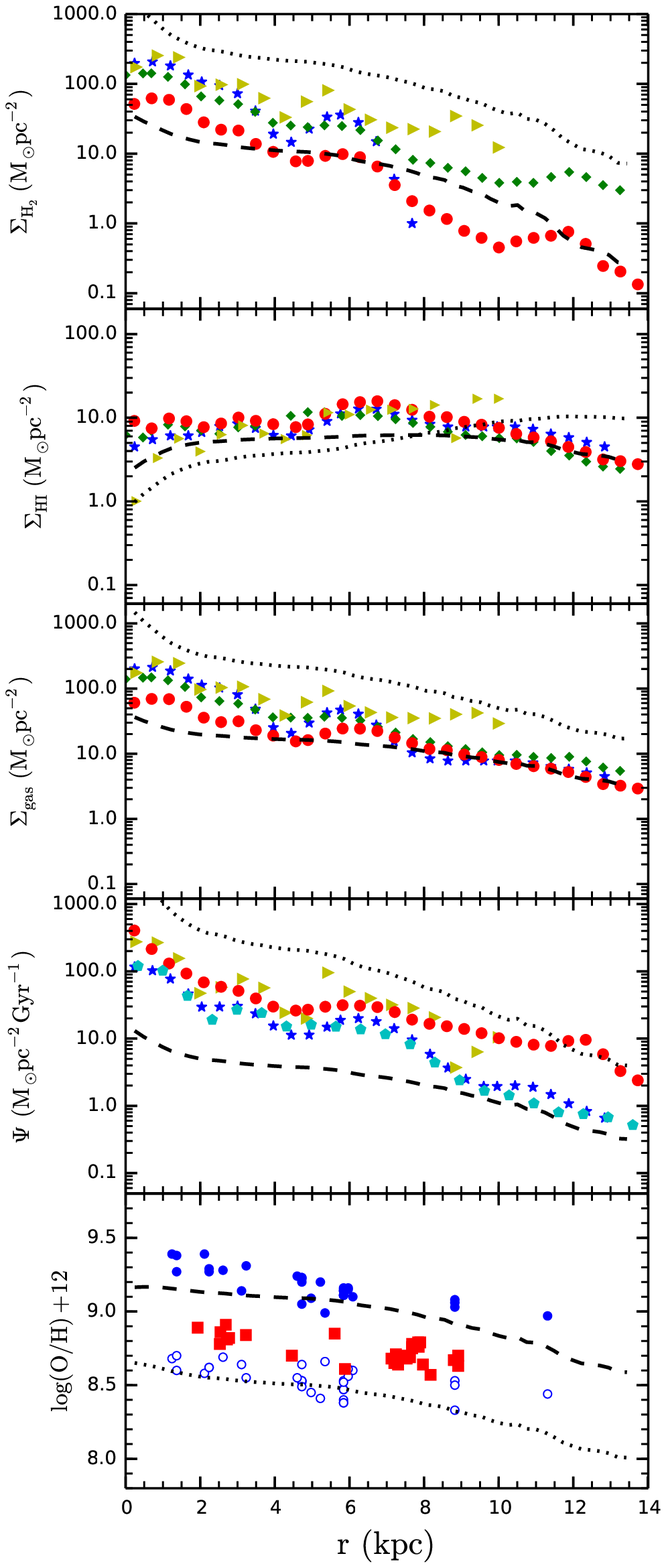}
   \caption{Comparisons of the radial profiles between model predictions
   and observations. The dashed and dotted lines are corresponding
   to the model results adopting $t_{\rm p}\,=\,0.1\,{\rm Gyr}$ and
   $t_{\rm p}\,=\,15\,{\rm Gyr}$, respectively. In the left-hand panels,
   the open circles show the observed profiles only
   corrected for the Galactic extinction and the filled
   ones also include a correction for the radial variation of internal
   extinction. The observed data of radial distributions of molecular,
   atomic, and total gas mass surface density shown in panels
   of the right-hand side are taken from \citet{schuster07} (filled circles),
   \citet{kennicutt07} (filled triangles), \citet{leroy08}
   (asterisks) and \citet{miyamoto14} (filled diamonds), and the SFR surface
   density obtained from \citet{schuster07} (filled circles), \citet{heesen14}
   (filled pentagons), \citet{leroy08} (asterisks) and \citet{kennicutt07}
   (filled triangles) are displayed in the right-fourth panel. The observed oxygen
   abundances taken from \citet{P14} are shown as filled squares,
   while the data taken from \citet{moustakas10} are plotted as open and filled
   circles corresponding to be derived by the PT05 and KK04 calibration method,
   respectively. }
   \label{Fig:results1}
   \end{figure*}
We also take into account the contribution of the gas outflow
process in our model. According to the mass-dependent
model of \citet{chang10}, we assume that the gas outflow rate
is proportional to the SFR and the coefficient is set to be
$b_{\rm out}\,=\,0.004$. It should be emphasized that our
final results is not sensitive to the variation of $b_{\rm out}$
since the outflow process plays a relatively small role in
the chemical evolution of such a massive galaxy as M51a
(the stellar mass of M51a is about
$M_{*}\,=\,10^{10.7}\,{\rm M_{\odot}}$).

We adopt the updated SPS model of \citet{BC03} (i.e., CB07) with the stellar
initial mass function (IMF) being taken from \citet{Chabrier03} in
our work. The lower and upper mass limits are adopted to be
$0.1\,{\rm M}_{\odot}$ and $100\,{\rm M}_{\odot}$, respectively.

Regarding the chemical evolution of the disc of M51a, both the
instantaneous-recycling approximation and the instantaneous mixing
of the gas with ejecta are assumed, that is, the gas in a fixed ring is
characterized by a unique composition at each epoch of time. We take
the classical set of equations of galactic chemical evolution from
\citet{Tinsley80}:
\begin{equation}
\frac{d[\Sigma_{\rm tot}(r,t)]}{dt}\,=\,f_{\rm{in}}(r,t)-f_{\rm{out}}(r,t),\\
\label{eq:tot}
\end{equation}
\begin{equation}
\frac{d[\Sigma_{\rm gas}(r,t)]}{dt}\,=\,-(1-R)\Psi(r,t)+f_{\rm{in}}(r,t)-f_{\rm{out}}(r,t),\\
\label{eq:gas}
\end{equation}
\begin{eqnarray}
\frac{d[Z(r,t)\Sigma_{\rm gas}(r,t)]}{dt}\,=\,y(1-R)\Psi(r,t)-Z(r,t)(1-R)\Psi(r,t) \nonumber\\
+Z_{\rm{in}}f_{\rm{in}}(r,t)-Z_{\rm{out}}(r,t)f_{\rm{out}}(r,t).
\label{eq:metallicity}
\end{eqnarray}
where $\Sigma_{\rm tot}(r,t)$ is the total (star + gas) mass surface density.
$Z(r,t)$ is the metallicity in the ring centered at
galactocentric distance $r$ at evolution time $t$. $R$ is the return
fraction and we set $R=0.3$
according to the adopted IMF. $y$ is the stellar yield and we set
$y=1\,{\rm Z}_{\odot}$ \citep{chang10,kang12}. $Z_{\rm{in}}$ is the metallicity
of the infalling gas and we assume the infalling gas is primordial,
that is $Z_{\rm{in}}=0$. $Z_{\rm{out}}(r,t)$ is the metallicity
of the outflowing gas and we assume that the outflow gas
has the same metallicity as that of ISM, e.g.,
$Z_{\rm{out}}(r,t)=Z(r,t)$ \citep{chang10,kang12}.

We emphasize that, under the condition that we have already
fixed $\sigma$ and A(r), there is only one free parameter left
in our model, the infall-peak time $t_{\rm p}$, which
regulates the shape of gas accretion history and then largely
influences the main properties of SFH along the disc
and then the evolution of M51a. Generally
speaking, $t_{\rm p}\,\rightarrow\,0$ is corresponding
to a time-declining infall rate, while
$t_{\rm p}\,\rightarrow\,\infty$ is corresponding to
a time-increasing gas infall rate. By comparing model
predictions with the observed data, we can discuss
the probable range of the free parameter in the model
and then know more about the main properties
of the evolution and SFH of M51a.

\section{Model Results Versus Observations}
\label{sect:result}

In this section, we present our results step by step.
Firstly, we investigate the influence of the free parameter
on our model predictions and present the viable model of
the evolution of M51a. Furthermore, we explore the
properties of stellar population along the disc and compare
the growth history of M51a with that of UGC\,8802 and the
Milky Way.

\subsection{Radial profiles}

   \begin{figure*}
   \centering
   \includegraphics[angle=0,height=18cm,width=11cm]{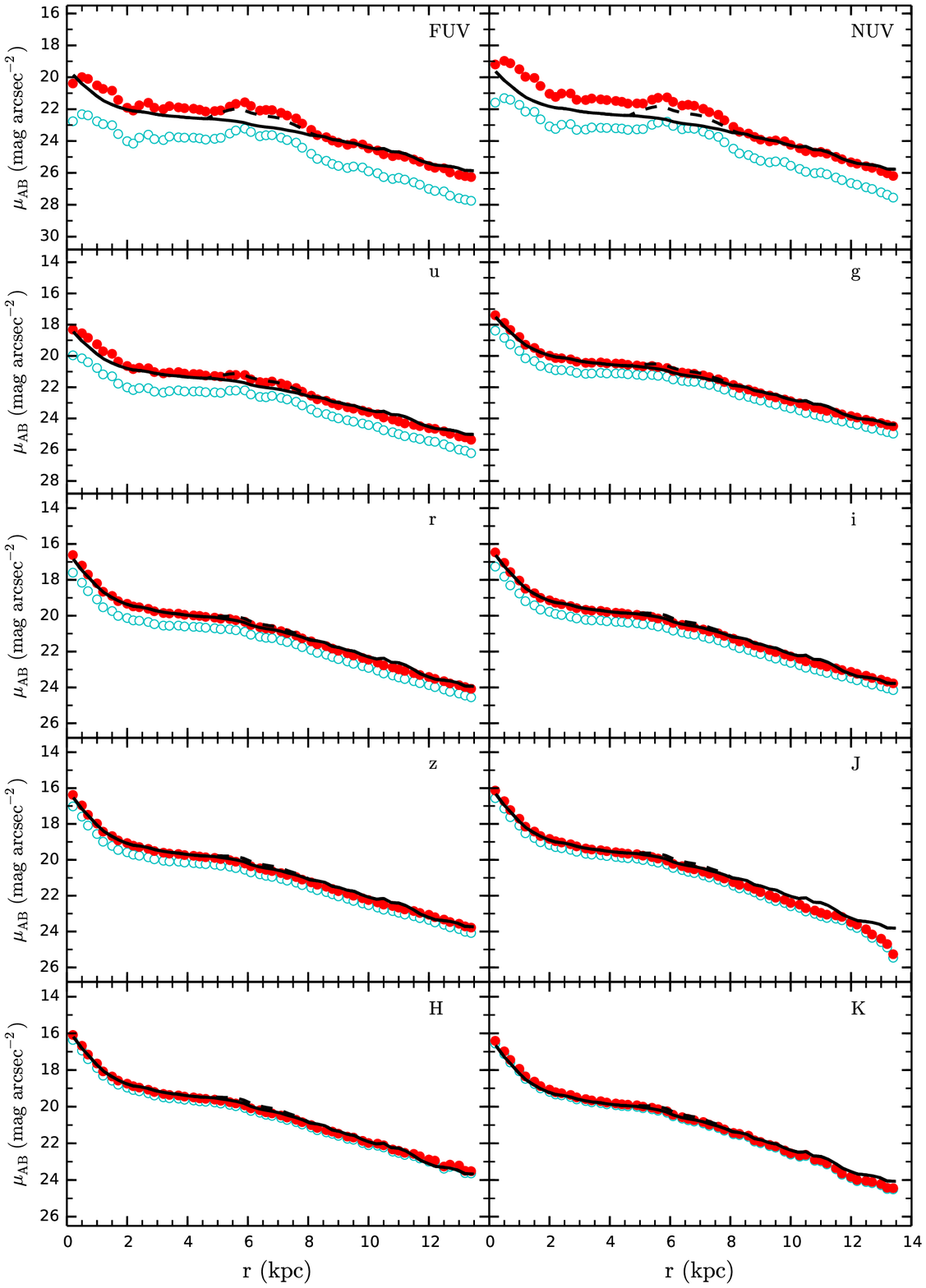}
   \includegraphics[angle=0,height=18cm,width=6cm]{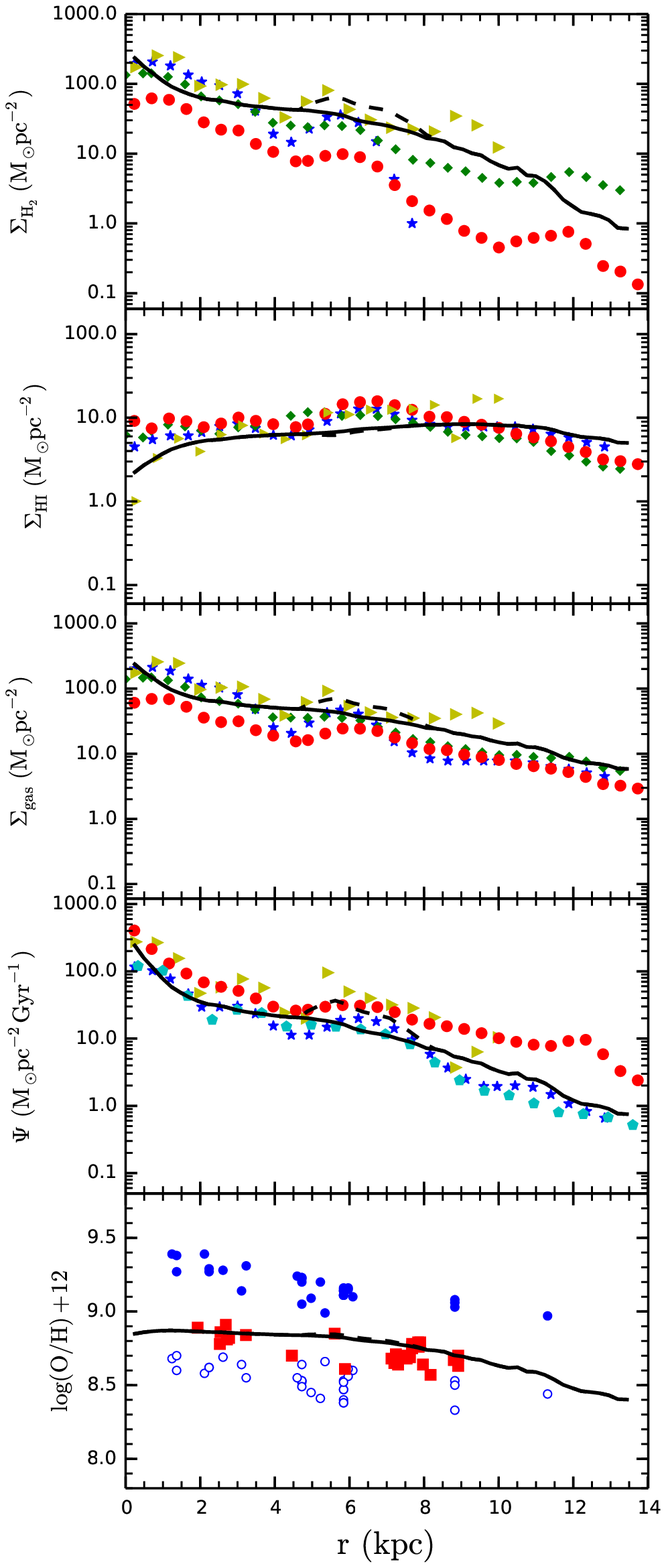}
   \caption{Comparisons of the model predictions with the observations.
    The solid lines plot the viable model results, while the dashed lines
   represent the 'toy' model results. The notation of the observational data is the same
   as that of Fig. \ref{Fig:results1}.}
   \label{Fig:results2}
   \end{figure*}

\begin{table}
\caption{The model predictions of the total quantities of M51a
with different parameters}
\label{Tab:predictions}
\begin{center}
\begin{tabular}{llllll}
\hline
\hline
$t_{\rm p}$  &   $M_{\rm K}$  &  $M_{\rm {H_2}}$  &  $M_{\rm {H_I}}$     &   SFR   & $f_{\rm gas}$\\
(Gyr)  &  (mag)  &  $(10^{9}\,{\rm M_{\odot}})$  &  $(10^{9}\,{\rm M_{\odot}})$  &  (${\rm M_{\odot}\,yr^{-1}}$) &  \\
\hline
$0.1$   &  -24.17      &  1.86   &   2.16   &  1.04   &   0.07   \\
$7.0$   &  -24.84      &  6.92   &   2.96   &  5.02   &   0.17   \\
$15.0$  &  -26.08      &  38.22  &   3.21   &  49.75  &   0.46    \\
\hline
\end{tabular}\\
\end{center}
\end{table}

To explore the influence of free parameter on model results,
we firstly consider two limiting cases of
$t_{\rm p}\,=\,0.1\,\rm Gyr$ and $t_{\rm p}\,=\,15\,\rm Gyr$
and present the comparison between model predictions of
the radial profiles and the observations in Fig. \ref{Fig:results1}.
The dashed and dotted lines in Fig. \ref{Fig:results1} denote the
predictions of the models adopting $t_{\rm p}\,=\,0.1\,\rm Gyr$
and $t_{\rm p}\,=\,15\,\rm Gyr$, respectively.
The left-hand side of Fig. \ref{Fig:results1} displays the radial
multi-bands surface brightness profiles from FUV-band to
$K$-band, where the open circles show the observed profiles only
corrected for the Galactic extinction and the filled
ones also include a correction for the radial variation of internal
extinction. The right-hand side of Fig. \ref{Fig:results1}
shows the $\rm H_{2}$, HI and total gas mass surface density, SFR
surface density, and oxygen abundance radial profiles.
The details of the observed data are presented in Section \ref{sect:Obs}.

Large differences between dashed and dotted lines in
Fig. \ref{Fig:results1} shows that the model predictions
are very sensitive to the adopted $t_{\rm p}$.
The case of $t_{\rm p}\,=\,0.1\,\rm Gyr$
(dashed lines) corresponds to a time-decreasing
gas-infall rate that most of the gas has been accreted to the
disc in the early period of its history, while that of
$t_{\rm p}\,=\,15\,\rm Gyr$ (dotted lines) corresponds
to a time-increasing gas infall-rate that a large fraction
of cold gas is still infalling to the disc at the present
time. It can be seen that the model adopting the earlier infall-peak time
(dashed lines) predicts less luminous surface brightness,
lower molecular gas surface density, lower SFR surface density
and higher gas-phase oxygen abundance than that adopting
a later infall-peak time (dotted lines). This is mainly due to
the fact that, in our model, the setting of
earlier infall-peak time corresponds to a faster gas accretion
and higher SF process in the earlier stage of its history and
then leads to the older stellar population age, higher
gas-phase metallicity and lower cold gas content at the present day.


Interestingly, we can see from Fig. \ref{Fig:results1} that
the area between the dashed and dotted lines brackets almost
the whole region of the observations, which indicates that
it is possible to construct a model that can reproduce the main
features of the observations of the M51a disc.
In fact, these two extreme cases ($t_{\rm p}\,=\,0.1\,\rm Gyr$ and
$t_{\rm p}\,=\,15\,\rm Gyr$) are bracketing the possible range
of our model results, from where we can see the variation trends of
the model predictions with different value of $t_{\rm p}$. After a
series of calculations and comparisons, we obtain a
viable model by adopting the infall-peak time
$t_{\rm p}/{\rm Gyr}\,=\,7.0$
and plot its results as solid lines in Fig.\ref{Fig:results2}.
The notation of observed data is the same as that of Fig.\ref{Fig:results1}.
We can see from Fig. \ref{Fig:results2} that the solid lines
can basically reproduce most of the observational data,
which indicates that our viable model includes and describes
reasonably the key ingredients of the main processes
that regulate the formation and evolution of M51a.

To further understand the effect of $t_{\rm p}$ on
the evolution of M51a, Table \ref{Tab:predictions} summarizes
the model predictions of the present total quantities of M51a,
including the absolute K-band magnitude $M_{\rm K}$,
the molecular gas mass $M_{\rm {H_2}}$, the atomic gas mass
$M_{\rm {H_I}}$, SFR and the gas fraction $f_{\rm gas}$.
Three different values of $t_{\rm p}$ ($0.1, 7.0, 15\rm Gyr$)
are considered, which corresponds to the dashed, dotted lines
in Fig. \ref{Fig:results1} and solid lines in Fig.\ref{Fig:results2}.
It can be seen from Table \ref{Tab:predictions} that the model
adopting a later infall-peak time predicts higher SFR and
higher gas fraction. The predictions of the viable model
adopting $t_{\rm p}/{\rm Gyr}\,=\,7.0$ are in fairy
agreement with the corresponding physical properties in Table
\ref{Tab:obs} considering the observed uncertainties.

However, it can be seen from Fig.\ref{Fig:results2} that,
contrary to the smoothness of the model predictions, the
observed radial profiles of cold gas and SFR surface density
show peaks at $R_{\rm G}\sim\,6\,\rm kpc$ and the
location of these peaks coincide with the location of
the most active SF region in the disc of M51a \citep{scheepmaker09}.
In addition, bumps obviously appear on the surface
brightness radial profiles in the FUV-, NUV- and $u$-band,
and the intensity of the bump decreases gradually with
the increase of the wavelength, which suggests that
active SF process may take place in recent years around there.

Indeed, both kinematic and hydrodynamic modeling suggest
that single or multiple encounters between these two
galaxies are occurred 300-500 Myr ago \citep{S&L00,Theis03,dobbs10}.
Studies of stellar populations of M51a also revealed that
a burst of SF occurred $340-500\,{\rm Myr}$ ago \citep{cooper12,tikhonov09}.
Moreover, \citet{miyamoto14} found that the local maximum
at $R_{\rm G}\sim\,6\,\rm kpc$ on the radial profiles of cold
gas in M51a could be caused by the kinks or fractures of
the spiral arms by the interaction with its companion galaxy.

\begin{figure}
  \centering
  \includegraphics[angle=0,width=0.475\textwidth]{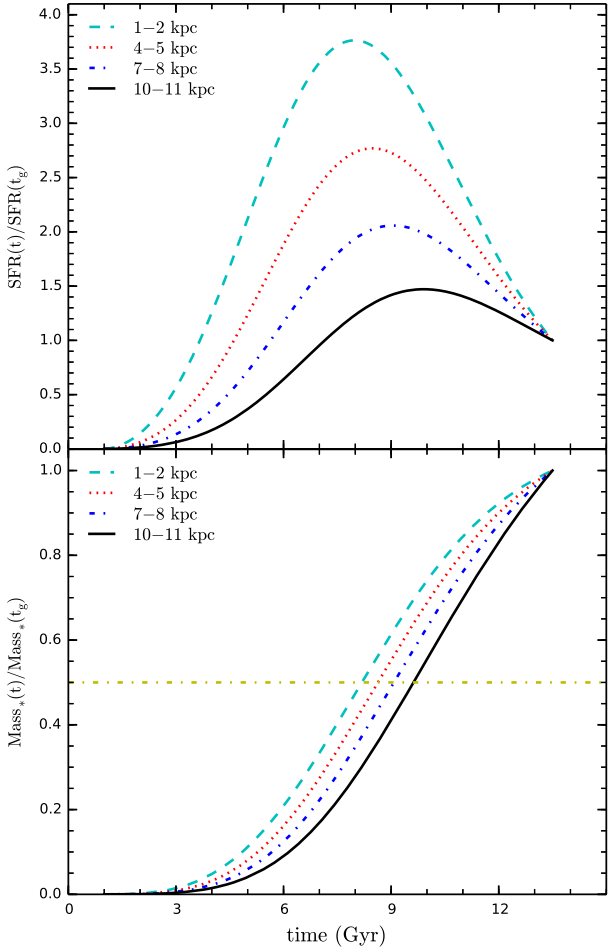}
    \caption{Evolution history of the M51a disc predicted by the viable
    model. Upper panel: the model predicted time evolution of SFR
    for four main spatial components (1-2\,kpc, 4-5\,kpc, 7-8\,kpc,
    10-11\,kpc) for M51a disc. Lower panel: relative stellar mass
    growth of the corresponding four spatial components as the upper panel.
    Both the SFRs and stellar masses are
    normalized by their present-day values. The horizontal
    dash-dotted line in the lower panel remarks when each component
    achieves 50\% of its final value.
    }
  \label{Fig:evolution}
\end{figure}

Correspondingly, we present a 'toy' model to imitate
the influence of the interaction between M51a and its companion
by allowing an additional cold gas infall (with constant rate
$8.31~{\rm M_{\odot}yr^{-1}}$) to the disc occurred
recently (during the period $13.0\leq\,t/{\rm Gyr}\,\leq13.2$)
round the region $R_{\rm G}\sim\,6\,\rm kpc$. Other ingredients
of the model are the same as those of the viable model.
The results of the 'toy' model are shown by dashed
lines in Fig. \ref{Fig:results2}. It can be seen that
the freshly infalling cold gas increases the mass
surface density of both the molecular gas and the SFR,
while has little effect on the
atomic surface density and the gas-phase
oxygen abundance. Moreover, the additional gas infall
results in more luminous surface brightness in FUV-, NUV-
and u-band, while the surface brightness in optical
and near infrared bands are less effected. Our results
suggest that the molecular gas surface density,
the SFR and the UV-band luminosity are important quantities
to explore the effects of recent interaction on galactic SF process.

\subsection{Stellar populations along the disc}

\begin{figure}
  \centering
  \includegraphics[angle=0,width=0.475\textwidth]{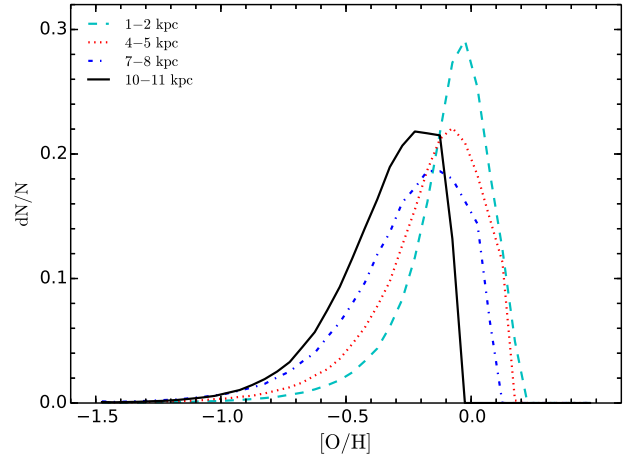}
    \caption{The viable model predicted metallicity distribution function for
    four main spatial regions (1-2\,kpc, 4-5\,kpc, 7-8\,kpc, and 10-11\,kpc)
    of M51a disc.
    }
  \label{Fig:Zdis}
\end{figure}

To further explore the properties of stellar population
along the disc of M51a, we select
four regions out and plot their viable model predictions of
the SFHs and growth curves of stellar masses in
Fig. \ref{Fig:evolution}, including 1-2\,kpc (dashed line),
4-5\,kpc (dotted line), 7-8\,kpc (dash-dotted line), 10-11\,kpc (solid line).
Correspondingly, Fig. \ref{Fig:Zdis} displays their
metallicity distribution function (MDF) predicted by
the viable model. Note that both the SFRs and stellar
masses in Fig. \ref{Fig:evolution} are normalized by
their present-day values, and the horizontal
dash-dotted line in the lower panel of Fig. \ref{Fig:evolution}
denotes the position that each component achieves 50\% of
its final value.

Fig. \ref{Fig:evolution} shows that the SFRs in the whole
disc are very low at early stage and increase
gradually due to the increase of gas infall rate. When
the consumption rate of cold gas to form new stars is roughly
balanced by the gas infall rate, the SFR reaches its
peak and then drops down to its present-day value.
It is interesting to see that the peak of the SFH moves
gradually to later time from the inner to outer parts
of disc. Moreover, Fig. \ref{Fig:Zdis} reveals that the inner
disc contains a higher percentage of metal-rich stars than
the outer disc. In other words, our model predictions
show evidences that the disc forms inside-out.
We will discuss this point again in next subsection.

\subsection{Comparisons with the Milky Way and UGC\,8802}

\begin{table*}
\caption{A summary of the viable models for M51a, UGC\,8802 and the Milky Way.}
\label{Tab:compare}
 \centering
 \begin{minipage}{175mm}
\begin{tabular}{llll}
\hline
\hline
Individual    &     M51a    &    UGC\,8802    &    Milky Way     \\
\hline
Input properties   &  $\mu_{\rm K}$\,$^{(a)}$     &  $\Sigma_*(r,t_{\rm g})=93.7\,{\rm e}^{-r/{\rm r_{d,u}}}\,^{(c)}$   &
$\Sigma_{\rm tot}(r,t_{\rm g})=55.0\,{\rm e}^{-(r-{\rm r_{\odot}})/{\rm r_{\rm d,mw}}}\,^{(d)}$   \\
&     $\Upsilon_{\star}^{\rm K}\,^{(b)}=\,0.5\,{\rm M_{\odot}}/{\rm L_{\odot,K}}$  &
${\rm r_{\rm d,u}}=5.8\,{\rm kpc}$ & ${\rm r_{\odot}}=8.5\,{\rm kpc}$ and ${\rm r_{\rm d,mw}}=2.7\,{\rm kpc}$\\
  \\
SF law   & $\Psi(r,t)=0.048\,\Sigma_{\rm{gas}}(r,t)^{1.56}$    &  $\Psi(r,t)=\Sigma_{\rm mol}(r,t)/0.77$  &
 $\Psi(r,t)=0.315\,\Sigma_{\rm{gas}}(r,t)^{1.4}(r/r_{\rm d,mw})^{-1}$   \\
  \\
Infal-peak time $t_{\rm p}$\,(Gyr)  &  7.0  &  $1.5\,r/{\rm r_{\rm d,u}}+5.0$   &  $2.0\,r/{\rm r_{\rm d,MW}}+2.0$   \\
\hline
\end{tabular}
Notes.\\
$^{(a)}$ The observed K-band surface brightness profile taken from \citet{MM091}.\\
$^{(b)}$ The $K$-band mass-to-light ratio taken from \citet{bell03} and \citet{leroy08}.\\
$^{(c)}$ ${\rm r_{\rm d,u}}$ is the disc scale-length of UGC\,8802 \citep{chang12}.\\
$^{(d)}$ ${\rm r_{\odot}}$ is the solar Galactic radius, and
${\rm r_{\rm d,mw}}$ is the disc scale-length of the Milky Way from \citet{chang99}.

\end{minipage}
\end{table*}

\begin{figure*}
   \centering
   \includegraphics[angle=0,width=0.85\textwidth]{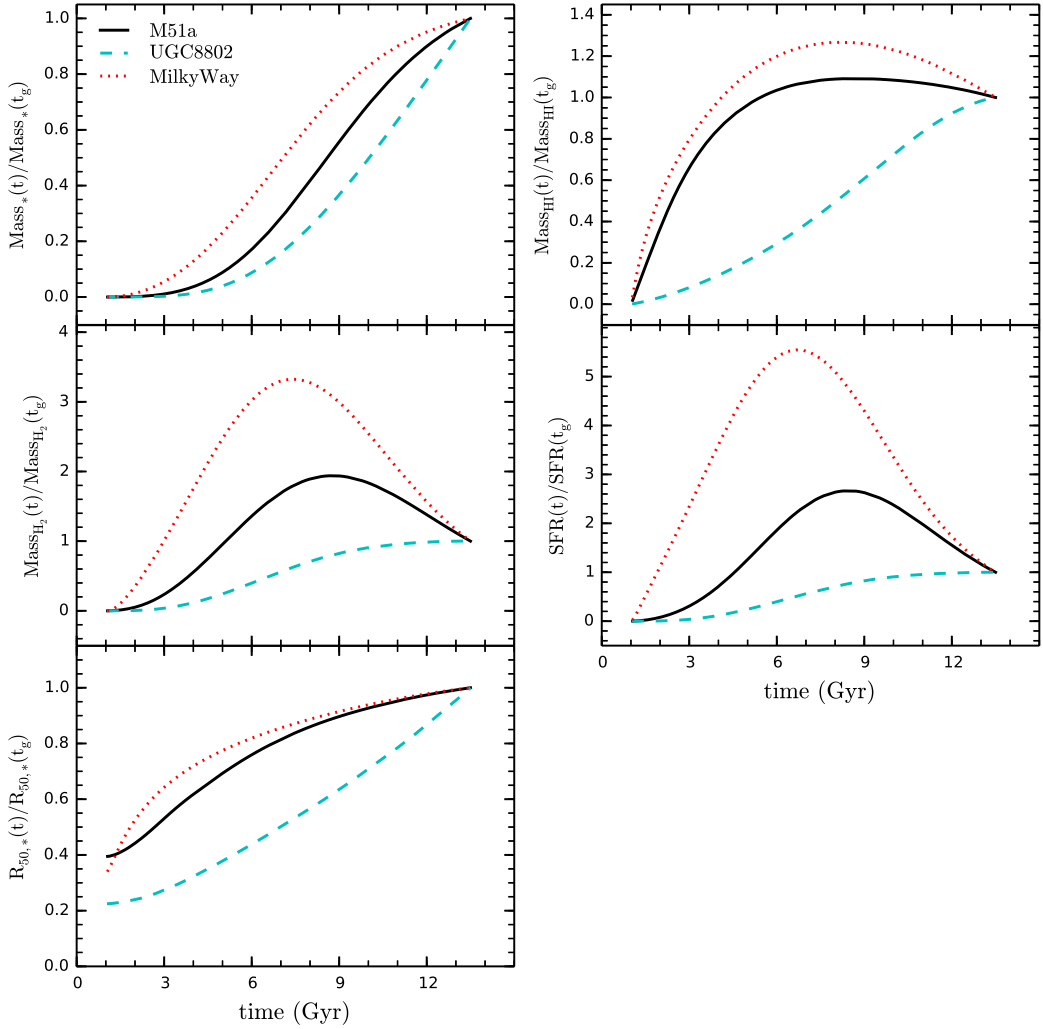}\\
   \caption{The growth history of M51a, UGC\,8802 and the Milky Way predicted by
   their individual viable models. The upper panels plot the model predicted time
   evolution of $M_{\rm *}$ (left panel) and $M_{\rm H_I}$ (right panel)
   of the three galaxies, The growth history of $M_{\rm H_2}$
   and the SFHS are displayed in the middle-left and middle-right panels,
   respectively. The model predicted evolution of the half mass size
   $R_{50,*}$ is displayed in the bottom-left panel. Each quantity is
   normalized to its value at the present day. The different line types
   correspond to different galaxies: solid lines for M51a, dashed lines
   for UGC\,8802, and dotted lines for the Milky Way.}
   \label{Fig:growth_history}
\end{figure*}

The Milky Way and UGC\,8802 are two disc galaxies containing
comparable stellar mass with M51a, and we have already
investigated SFHs of the two galaxies in previous
works \citep{chang99,chang12}. In this section, we will
compare the growth history of M51a predicted by the
viable model with that of the Milky Way and UGC\,8802.
We referred to \citet{chang99,chang12} for a
more indepth description of the models. Here we only
repeat main ingredients of the models such as the gas
infall rate and the SF law, and present a summary of
the models in Table \ref{Tab:compare}.

In our previous models of the Milky Way and
UGC\,8802, an Gaussian formula of gas
infall rate has been adopted. The input condition
of UGC\,8802 to constrain how much gas infalling to
the disc is the stellar mass surface density at the
present day, which is assumed to be pure exponential
$\Sigma_{*}(r,{\rm t_g})=93.7\,{\rm e}^{-r/{\rm r_{\rm d,u}}}$, with
radial scale-length ${\rm r_{\rm d,u}=5.8\,kpc}$ \citep{chang12}. In the case of
the Milky Way, the input parameter to constrain
the gas infall rate is the total (gas+star) mass
surface density at the present day, i.e.,
$\Sigma_{\rm tot}(r,{\rm t_g})=55.0\,{\rm e}^{-(r-r_{\odot})/{\rm r_{\rm d,mw}}}$,
where the solar Galactic radius is set to be
${\rm r_{\odot}=8.5\,kpc}$ and the disc scale-length is
${\rm r_{\rm d,mw}=2.7\,kpc}$ \citep{chang99}. Regarding the
SF law, the model of UGC\, 8802 adopted the one
that the SFR is proportional to the molecular hydrogen
surface density $\Psi(r,t)=\Sigma_{\rm mol}(r,t)/0.77$,
while the model of the Milky Way adopted a
radial-dependent Schmidt SF law, i.e., $\Psi(r,t)=
0.85\,\Sigma_{\rm gas}^{1.4}(r,t)(r/{\rm r_{\rm d,mw}})^{-1}$,
to be consistent with
the observed steep oxygen gradient along the disc.

In Fig. \ref{Fig:growth_history}, we compared model
predictions of the growth history of M51a (solid lines),
UGC\,8802 (dashed lines) and the Milky Way (dotted lines).
The upper panels plot the evolution of stellar mass
$M_{\rm *}$ (left panel) and atomic hydrogen mass
$M_{\rm H_I}$ (right panel). The growth history of molecular
hydrogen mass $M_{\rm H_2}$ and the SFHs of these three
galaxies are displayed in the middle-left and middle-right panels,
respectively. The bottom left panel shows the evolution of
the half mass size $R_{50,*}$, which is defined as
the radius at which half of the total stellar mass
is contained. Each quantity in Fig. \ref{Fig:growth_history}
is normalized to its value at the present day.

From Fig. \ref{Fig:growth_history} we can see that, at early
stage of evolution, most of the infalling cold gas is in
the form of atomic gas. After atomic gas accumulates
significantly enough, molecular gas begins to form and
then the SFR speeds up. The middle panels of
Fig. \ref{Fig:growth_history} show that the shape
of the SFH is very similar to the evolution curve of
the molecular gas, no matter whether the adopted SF law
assumes the SFR proportional to the molecular
gas surface density or not. This point suggests that
the density of molecular gas may be the main driver of
SF process, but this question is far from settled.

The growth curves of the mass of atomic gas show that,
both M51a and the Milky Way reach their peaks around
$4 \sim 7$\,Gyr ago and then slowly drop down to
their present-day values, while the atomic mass of
UGC\,8802 increases almost steadily and seems to
just approach to its peak. Indeed, the outstanding
point of UGC\,8802 is its extremely high neutral
gas content compared to other disc galaxies in this
stellar mass range \citep{GA09,CS10}. According to the model of
\citet{chang12}, a late infall-peak time is
necessary to explain the high neutral gas content of
UGC\,8802, which implies that UGC\,8802 is a
young galaxy and still active in SF processes.
This point can also be seen from both the upper-left
panel and the middle-right panel of Fig. \ref{Fig:growth_history}.
It is shown that half of the total stellar
mass of M51a has been assembled during the last $\sim$5.0\,Gyr,
while that for the Milky Way and UGC\,8802 is around
$\sim$6.5\,Gyr and $\sim$4.0\,Gyr ago, respectively.
In other words, our results suggest that the stellar
population of M51a is younger than that of the Milky Way,
but older than that of the gas-rich disc galaxy UGC\, 8802.

We also compare the evolution of the half-mass size $R_{50,*}$
among these three galaxies in the bottom left panel of
Fig. \ref{Fig:growth_history}.
Since the models of the Milky Way and UGC\, 8802 both
assume that the infall-peak time increases with radius
(see Table \ref{Tab:compare}), it is natural to see that both the
dotted and dashed lines increase with time. In the case
of M51a, although the model adopt a constant
infall-peak time, the model predictions also support
the disc inside-out formation scenario. This trend
also reinforces the recent analysis of the SFH
of CALIFA galaxies, which shows that galaxies more massive
than $10^{10}\,\rm M_{\odot}$ grow inside-out \citep{perez13,Delgado14}.
In fact, this inside-out formation mechanism has already successfully
applied to previous models of formation and evolution of disc galaxies
\citep{chang99,chang12,bp00,hou00,chiappini01,yin09,kang12} and
supported by the observations for nearby disc galaxies
\citep{MM07, perez13}.

\section{Summary}
\label{sect:summary}

M51a is a grand-design spiral galaxy and well known to be
interacting with its companion. In this paper, we focus on
investigating the SFHs of M51a by constructing a
parameterized model and comparing model predictions with
the observed data, especially the radial distributions
of cold gas surface density, metallicity, and the
radial profiles of surface brightness in multi-bands.
Our main results can be summarized as follows:

\begin{enumerate}

\item Our results show that the model predictions are very
sensitive to the adopted infall-peak time $t_{\rm p}$. A late
infall-peak time $t_{\rm p}$ results in more luminous
surface brightness, low metallicity, high gas
and SFR surface densities.

\item The model adopting a constant infall-peak time
$t_{\rm p}\,=\,7.0{\rm Gyr}$ can nicely reproduce
most of the observed constraints of M51a. We also
introduce a 'toy' model, which allows a small amount
of additional cold gas infall to the disc in recent
time, to imitate the influence of the interaction
between M51a and its companion. Our results show
that the additional gas infall could be attributed to the
observed small bumps on the radial profiles
of cold gas surface density, SFR surface density and
surface brightness in the UV-band.

\item Although we adopt a constant infall-peak
time, our model predictions still show evidence that
the disc forms inside-out. This is consistent with
the results of \citet{perez13} and \citet{Delgado14} that
galaxies more massive than $10^{10}\,\rm M_{\odot}$
grow inside-out.

\item We also compared the model predicted growth
histories of M51a, the Milky Way and UGC\,8802. We
find that the mean stellar age of M51a is younger than
that of the Milky Way, but older than that of the
gas-rich disc galaxy UGC\,8802. Our results also show that
half of the total stellar mass of M51a may has been assembled
during the last $\sim$5.0\,Gyr and SF may still
proceed actively in the disc
of M51a.

We emphasize that here we only present a parameterized model. Through
the comparison between model predictions and observations,
our ultimate goal is to present our story of the main properties
of the formation and evolution of M51a. Although the accurate value
of $t_{\rm p}$ in the viable model is not unique, the main
conclusions of our results are robust. However, since M51a
belongs to a dynamically complicted system and maintains
grand-designed spiral structure, it is a long way to go
to clearly understand its origin and evolution.



\end{enumerate}

\section*{acknowledgements}

We thank the referee for thoughtful comments and insightful
suggestions that improved our paper greatly.
Xiaoyu Kang and Fenghui Zhang are supported by the National
Natural Science Foundation of China (NSFC)
grant No. 11403092, 11273053, 11033008, 11373063, and Yunnan
Foundation No. 2011CI053.
Ruixiang Chang is supported by the NSFC grant No.11373053, 11390373,
and Strategic Priority Research Program "The Emergence of
Cosmological Structures" of the Chinese Academy of Sciences
(CAS; grant XDB09010100).

\bibliography{kxy}

\begin{thebibliography}{}

\bibitem[\protect\citeauthoryear{{Bell}, {McIntosh}, {Katz} \&
  {Weinberg}}{{Bell} et~al.}{2003}]{bell03}
{Bell} E.~F.,  {McIntosh} D.~H.,  {Katz} N.,    {Weinberg} M.~D.,  2003, \apjs,
  149, 289

\bibitem[\protect\citeauthoryear{{Bigiel}, {Leroy}, {Walter}, {Brinks}, {de
  Blok}, {Madore} \& {Thornley}}{{Bigiel} et~al.}{2008}]{bigiel08}
{Bigiel} F.,  {Leroy} A.,  {Walter} F.,  {Brinks} E.,  {de Blok} W.~J.~G.,
  {Madore} B.,    {Thornley} M.~D.,  2008, \aj, 136, 2846

\bibitem[\protect\citeauthoryear{{Blitz} \& {Rosolowsky}}{{Blitz} \&
  {Rosolowsky}}{2006}]{blitz06}
{Blitz} L.,  {Rosolowsky} E.,  2006, \apj, 650, 933

\bibitem[\protect\citeauthoryear{{Boissier} \& {Prantzos}}{{Boissier} \&
  {Prantzos}}{2000}]{bp00}
{Boissier} S.,  {Prantzos} N.,  2000, \mnras, 312, 398

\bibitem[\protect\citeauthoryear{{Boselli}, {Gavazzi} \& {Sanvito}}{{Boselli}
  et~al.}{2003}]{boselli03}
{Boselli} A.,  {Gavazzi} G.,    {Sanvito} G.,  2003, \aap, 402, 37

\bibitem[\protect\citeauthoryear{{Bresolin}, {Garnett} \& {Kennicutt}
  Jr.}{{Bresolin} et~al.}{2004}]{bresolin04}
{Bresolin} F.,  {Garnett} D.~R.,    {Kennicutt} Jr. R.~C.,  2004, \apj, 615,
  228

\bibitem[\protect\citeauthoryear{{Bruzual} \& {Charlot}}{{Bruzual} \&
  {Charlot}}{2003}]{BC03}
{Bruzual} G.,  {Charlot} S.,  2003, \mnras, 344, 1000

\bibitem[\protect\citeauthoryear{{Catinella}, {Schiminovich}, {Kauffmann},
  {Fabello}, {Wang}, {Hummels}, {Lemonias}, {Wild} \& {Wyder}}{{Catinella}
  et~al.}{2010}]{CS10}
{Catinella} B.,  {Schiminovich} D.,  {Kauffmann} G.,  {Fabello} S.,  {Wang} J.,
   {Hummels} C.,  {Lemonias} J.,  {Wild} V.,    {Wyder} T.~K.,  2010, \mnras,
  403, 683

\bibitem[\protect\citeauthoryear{{Chabrier}}{{Chabrier}}{2003}]{Chabrier03}
{Chabrier} G.,  2003, \apjl, 586, L133

\bibitem[\protect\citeauthoryear{{Chang}, {Hou}, {Shen} \& {Shu}}{{Chang}
  et~al.}{2010}]{chang10}
{Chang} R.~X.,  {Hou} J.~L.,  {Shen} S.~Y.,    {Shu} C.~G.,  2010, \apj, 722,
  380

\bibitem[\protect\citeauthoryear{{Chang}, {Hou}, {Shu} \& {Fu}}{{Chang}
  et~al.}{1999}]{chang99}
{Chang} R.~X.,  {Hou} J.~L.,  {Shu} C.~G.,    {Fu} C.~Q.,  1999, \aap, 350, 38

\bibitem[\protect\citeauthoryear{{Chang}, {Shen} \& {Hou}}{{Chang}
  et~al.}{2012}]{chang12}
{Chang} R.~X.,  {Shen} S.~Y.,    {Hou} J.~L.,  2012, \apjl, 753, L10

\bibitem[\protect\citeauthoryear{{Chiappini}, {Matteucci} \&
  {Romano}}{{Chiappini} et~al.}{2001}]{chiappini01}
{Chiappini} C.,  {Matteucci} F.,    {Romano} D.,  2001, \apj, 554, 1044

\bibitem[\protect\citeauthoryear{{Colombo}, {Meidt}, {Schinnerer},
  {Garc{\'{\i}}a-Burillo}, {Hughes}, {Pety}, {Leroy}, {Dobbs}, {Dumas},
  {Thompson}, {Schuster} \& {Kramer}}{{Colombo} et~al.}{2014}]{Colombo14}
{Colombo} D.,  {Meidt} S.~E.,  {Schinnerer} E.,  {Garc{\'{\i}}a-Burillo} S.,
  {Hughes} A.,  {Pety} J.,  {Leroy} A.~K.,  {Dobbs} C.~L.,  {Dumas} G.,
  {Thompson} T.~A.,  {Schuster} K.~F.,    {Kramer} C.,  2014, \apj, 784, 4

\bibitem[\protect\citeauthoryear{{Cortese}, {Boselli}, {Franzetti}, {Decarli},
  {Gavazzi}, {Boissier} \& {Buat}}{{Cortese} et~al.}{2008}]{cortese08}
{Cortese} L.,  {Boselli} A.,  {Franzetti} P.,  {Decarli} R.,  {Gavazzi} G.,
  {Boissier} S.,    {Buat} V.,  2008, \mnras, 386, 1157

\bibitem[\protect\citeauthoryear{{Dalcanton}}{{Dalcanton}}{2007}]{dalcanton07}
{Dalcanton} J.~J.,  2007, \apj, 658, 941

\bibitem[\protect\citeauthoryear{{Dobbs}, {Theis}, {Pringle} \& {Bate}}{{Dobbs}
  et~al.}{2010}]{dobbs10}
{Dobbs} C.~L.,  {Theis} C.,  {Pringle} J.~E.,    {Bate} M.~R.,  2010, \mnras,
  403, 625

\bibitem[\protect\citeauthoryear{{Elmegreen}}{{Elmegreen}}{1989}]{elmegreen89_%
1}
{Elmegreen} B.~G.,  1989, \apj, 338, 178

\bibitem[\protect\citeauthoryear{{Garcia-Appadoo}, {West}, {Dalcanton},
  {Cortese} \& {Disney}}{{Garcia-Appadoo} et~al.}{2009}]{GA09}
{Garcia-Appadoo} D.~A.,  {West} A.~A.,  {Dalcanton} J.~J.,  {Cortese} L.,
  {Disney} M.~J.,  2009, \mnras, 394, 340

\bibitem[\protect\citeauthoryear{{Gonz{\'a}lez Delgado}, {P{\'e}rez}, {Cid
  Fernandes}, {Garc{\'{i}}a-Benito}, {de Amorim}, {S{\'a}nchez}, {Marino},
  {Quirrenbach}, {V{\'{\i}}lchez} \& {Wisotzki}}{{Gonz{\'a}lez Delgado}
  et~al.}{2014}]{Delgado14}
{Gonz{\'a}lez Delgado} R.~M.,  {P{\'e}rez} E.,  {Cid Fernandes} R.,
  {Garc{\'{i}}a-Benito} R.,  {de Amorim} A.~L.,  {S{\'a}nchez} S.~F.,  {Marino}
  R.~A.,  {Quirrenbach} A.,  {V{\'{\i}}lchez} J.~M.,    {Wisotzki} L.,  2014,
  \aap, 562, A47

\bibitem[\protect\citeauthoryear{{Heesen}, {Brinks}, {Leroy}, {Heald}, {Braun},
  {Bigiel} \& {Beck}}{{Heesen} et~al.}{2014}]{heesen14}
{Heesen} V.,  {Brinks} E.,  {Leroy} A.~K.,  {Heald} G.,  {Braun} R.,  {Bigiel}
  F.,    {Beck} R.,  2014, \aj, 147, 103

\bibitem[\protect\citeauthoryear{{Helfer}, {Thornley}, {Regan}, {Wong},
  {Sheth}, {Vogel}, {Blitz} \& {Bock}}{{Helfer} et~al.}{2003}]{helfer03}
{Helfer} T.~T.,  {Thornley} M.~D.,  {Regan} M.~W.,  {Wong} T.,  {Sheth} K.,
  {Vogel} S.~N.,  {Blitz} L.,    {Bock} D.~C.-J.,  2003, \apjs, 145, 259

\bibitem[\protect\citeauthoryear{{Hitschfeld}, {Kramer}, {Schuster},
  {Garcia-Burillo} \& {Stutzki}}{{Hitschfeld} et~al.}{2009}]{hitschfeld09}
{Hitschfeld} M.,  {Kramer} C.,  {Schuster} K.~F.,  {Garcia-Burillo} S.,
  {Stutzki} J.,  2009, \aap, 495, 795

\bibitem[\protect\citeauthoryear{{Hou}, {Prantzos} \& {Boissier}}{{Hou}
  et~al.}{2000}]{hou00}
{Hou} J.~L.,  {Prantzos} N.,    {Boissier} S.,  2000, \aap, 362, 921

\bibitem[\protect\citeauthoryear{{Hu}, {Shao} \& {Peng}}{{Hu}
  et~al.}{2013}]{hu13}
{Hu} T.,  {Shao} Z.,    {Peng} Q.,  2013, \apjl, 762, L27

\bibitem[\protect\citeauthoryear{{Hwang} \& {Lee}}{{Hwang} \&
  {Lee}}{2008}]{Hwang08}
{Hwang} N.,  {Lee} M.~G.,  2008, \aj, 135, 1567

\bibitem[\protect\citeauthoryear{{Hwang} \& {Lee}}{{Hwang} \&
  {Lee}}{2010}]{Hwang10}
{Hwang} N.,  {Lee} M.~G.,  2010, \apj, 709, 411

\bibitem[\protect\citeauthoryear{{Jarrett}, {Chester}, {Cutri}, {Schneider} \&
  {Huchra}}{{Jarrett} et~al.}{2003}]{jarrett03}
{Jarrett} T.~H.,  {Chester} T.,  {Cutri} R.,  {Schneider} S.~E.,    {Huchra}
  J.~P.,  2003, \aj, 125, 525

\bibitem[\protect\citeauthoryear{{Kaleida} \& {Scowen}}{{Kaleida} \&
  {Scowen}}{2010}]{Kaleida10}
{Kaleida} C.,  {Scowen} P.~A.,  2010, \aj, 140, 379

\bibitem[\protect\citeauthoryear{{Kang}, {Chang}, {Yin}, {Hou}, {Zhang},
  {Zhang} \& {Han}}{{Kang} et~al.}{2012}]{kang12}
{Kang} X.,  {Chang} R.,  {Yin} J.,  {Hou} J.,  {Zhang} F.,  {Zhang} Y.,
  {Han} Z.,  2012, \mnras, 426, 1455

\bibitem[\protect\citeauthoryear{{Kennicutt} Jr.}{{Kennicutt}}{1998}]{k98}
{Kennicutt} Jr. R.~C.,  1998, \araa, 36, 189

\bibitem[\protect\citeauthoryear{{Kennicutt} Jr., {Armus}, {Bendo}, {Calzetti},
  {Dale}, {Draine}, {Engelbracht}, {Rieke}, {Roussel}, {Smith}, {Thornley} \&
  {Walter}}{{Kennicutt} et~al.}{2003}]{kennicutt03}
{Kennicutt} Jr. R.~C.,  {Armus} L.,  {Bendo} G.,  {Calzetti} D.,  {Dale} D.~A.,
   {Draine} B.~T.,  {Engelbracht} C.~W.,  {Rieke} M.~J.,  {Roussel} H.,
  {Smith} J.-D.~T.,  {Thornley} M.~D.,    {Walter} F.,  2003, \pasp, 115, 928

\bibitem[\protect\citeauthoryear{{Kennicutt} Jr., {Calzetti}, {Walter},
  {Helou}, {Hollenbach}, {Armus}, {Bendo}, {de Mello}, {Meyer}, {Moustakas},
  {Murphy}, {Sheth} \& {Smith}}{{Kennicutt} et~al.}{2007}]{kennicutt07}
{Kennicutt} Jr. R.~C.,  {Calzetti} D.,  {Walter} F.,  {Helou} G.,  {Hollenbach}
  D.~J.,  {Armus} L.,  {Bendo} G.,  {de Mello} D.,  {Meyer} M.,  {Moustakas}
  J.,  {Murphy} E.~J.,  {Sheth} K.,    {Smith} J.~D.~T.,  2007, \apj, 671, 333

\bibitem[\protect\citeauthoryear{{Kobulnicky} \& {Kewley}}{{Kobulnicky} \&
  {Kewley}}{2004}]{KK04}
{Kobulnicky} H.~A.,  {Kewley} L.~J.,  2004, \apj, 617, 240

\bibitem[\protect\citeauthoryear{{Korotin}, {Andrievsky}, {Luck}, {L{\'e}pine},
  {Maciel} \& {Kovtyukh}}{{Korotin} et~al.}{2014}]{korotin14}
{Korotin} S.~A.,  {Andrievsky} S.~M.,  {Luck} R.~E.,  {L{\'e}pine} J.~R.~D.,
  {Maciel} W.~J.,    {Kovtyukh} V.~V.,  2014, \mnras, 444, 3301

\bibitem[\protect\citeauthoryear{{Lee}, {Kim}, {Park}, {Ree}, {Kyeong} \&
  {Chung}}{{Lee} et~al.}{2011}]{lee11}
{Lee} J.~H.,  {Kim} S.~C.,  {Park} H.~S.,  {Ree} C.~H.,  {Kyeong} J.,
  {Chung} J.,  2011, \apj, 740, 42

\bibitem[\protect\citeauthoryear{{Lee}, {Chandar} \& {Whitmore}}{{Lee}
  et~al.}{2005}]{lee05}
{Lee} M.~G.,  {Chandar} R.,    {Whitmore} B.~C.,  2005, \aj, 130, 2128

\bibitem[\protect\citeauthoryear{{Leroy}, {Walter}, {Brinks}, {Bigiel}, {de
  Blok}, {Madore} \& {Thornley}}{{Leroy} et~al.}{2008}]{leroy08}
{Leroy} A.~K.,  {Walter} F.,  {Brinks} E.,  {Bigiel} F.,  {de Blok} W.~J.~G.,
  {Madore} B.,    {Thornley} M.~D.,  2008, \aj, 136, 2782

\bibitem[\protect\citeauthoryear{{Lin} \& {Shu}}{{Lin} \&
  {Shu}}{1964}]{lin&shu}
{Lin} C.~C.,  {Shu} F.~H.,  1964, \apj, 140, 646

\bibitem[\protect\citeauthoryear{{Lin}, {Zou}, {Kong}, {Lin}, {Mao}, {Cheng},
  {Jiang} \& {Zhou}}{{Lin} et~al.}{2013}]{linl13}
{Lin} L.,  {Zou} H.,  {Kong} X.,  {Lin} X.,  {Mao} Y.,  {Cheng} F.,  {Jiang}
  Z.,    {Zhou} X.,  2013, \apj, 769, 127

\bibitem[\protect\citeauthoryear{{Matteucci}, {Spitoni}, {Recchi} \&
  {Valiante}}{{Matteucci} et~al.}{2009}]{matteucci09}
{Matteucci} F.,  {Spitoni} E.,  {Recchi} S.,    {Valiante} R.,  2009, \aap,
  501, 531

\bibitem[\protect\citeauthoryear{{Mentuch Cooper}, {Wilson}, {Foyle}, {Bendo},
  {Koda}, {Baes}, {Boquien}, {Boselli}, {Ciesla}, {Cooray}, {Eales}, {Sauvage},
  {Spinoglio} \& {Smith}}{{Mentuch Cooper} et~al.}{2012}]{cooper12}
{Mentuch Cooper} E.,  {Wilson} C.~D.,  {Foyle} K.,  {Bendo} G.,  {Koda} J.,
  {Baes} M.,  {Boquien} M.,  {Boselli} A.,  {Ciesla} L.,  {Cooray} A.,  {Eales}
  S.,  {Sauvage} M.,  {Spinoglio} L.,    {Smith} M.~W.~L.,  2012, \apj, 755,
  165

\bibitem[\protect\citeauthoryear{{Miyamoto}, {Nakai} \& {Kuno}}{{Miyamoto}
  et~al.}{2014}]{miyamoto14}
{Miyamoto} Y.,  {Nakai} N.,    {Kuno} N.,  2014, \pasj, 66, 36

\bibitem[\protect\citeauthoryear{{Moll{\'a}} \& {D{\'{\i}}az}}{{Moll{\'a}} \&
  {D{\'{\i}}az}}{2005}]{m&d05}
{Moll{\'a}} M.,  {D{\'{\i}}az} A.~I.,  2005, \mnras, 358, 521

\bibitem[\protect\citeauthoryear{{Moustakas}, {Kennicutt} Jr., {Tremonti},
  {Dale}, {Smith} \& {Calzetti}}{{Moustakas} et~al.}{2010}]{moustakas10}
{Moustakas} J.,  {Kennicutt} Jr. R.~C.,  {Tremonti} C.~A.,  {Dale} D.~A.,
  {Smith} J.-D.~T.,    {Calzetti} D.,  2010, \apjs, 190, 233

\bibitem[\protect\citeauthoryear{{Mu{\~n}oz-Mateos}, {Boissier}, {Gil de Paz},
  {Zamorano}, {Kennicutt} Jr., {Moustakas}, {Prantzos} \&
  {Gallego}}{{Mu{\~n}oz-Mateos} et~al.}{2011}]{MM11}
{Mu{\~n}oz-Mateos} J.~C.,  {Boissier} S.,  {Gil de Paz} A.,  {Zamorano} J.,
  {Kennicutt} Jr. R.~C.,  {Moustakas} J.,  {Prantzos} N.,    {Gallego} J.,
  2011, \apj, 731, 10

\bibitem[\protect\citeauthoryear{{Mu{\~n}oz-Mateos}, {Gil de Paz}, {Boissier},
  {Zamorano}, {Jarrett}, {Gallego} \& {Madore}}{{Mu{\~n}oz-Mateos}
  et~al.}{2007}]{MM07}
{Mu{\~n}oz-Mateos} J.~C.,  {Gil de Paz} A.,  {Boissier} S.,  {Zamorano} J.,
  {Jarrett} T.,  {Gallego} J.,    {Madore} B.~F.,  2007, \apj, 658, 1006

\bibitem[\protect\citeauthoryear{{Mu{\~n}oz-Mateos}, {Gil de Paz}, {Zamorano},
  {Boissier}, {Dale}, {P{\'e}rez-Gonz{\'a}lez}, {Gallego}, {Madore}, {Bendo},
  {Boselli}, {Buat}, {Calzetti}, {Moustakas} \& {Kennicutt}
  Jr.}{{Mu{\~n}oz-Mateos} et~al.}{2009}]{MM091}
{Mu{\~n}oz-Mateos} J.~C.,  {Gil de Paz} A.,  {Zamorano} J.,  {Boissier} S.,
  {Dale} D.~A.,  {P{\'e}rez-Gonz{\'a}lez} P.~G.,  {Gallego} J.,  {Madore}
  B.~F.,  {Bendo} G.,  {Boselli} A.,  {Buat} V.,  {Calzetti} D.,  {Moustakas}
  J.,    {Kennicutt} Jr. R.~C.,  2009, \apj, 703, 1569

\bibitem[\protect\citeauthoryear{{Ostriker}, {McKee} \& {Leroy}}{{Ostriker}
  et~al.}{2010}]{Ostriker10}
{Ostriker} E.~C.,  {McKee} C.~F.,    {Leroy} A.~K.,  2010, \apj, 721, 975

\bibitem[\protect\citeauthoryear{{P{\'e}rez}, {Cid Fernandes}, {Gonz{\'a}lez
  Delgado}, {Garc{\'{i}}a-Benito}, {S{\'a}nchez}, {Husemann}, {de Amorim}, {van
  de Ven}, {Walcher}, {Wisotzki}, {Cortijo-Ferrero} \& {CALIFA
  Collaboration}}{{P{\'e}rez} et~al.}{2013}]{perez13}
{P{\'e}rez} E.,  {Cid Fernandes} R.,  {Gonz{\'a}lez Delgado} R.~M.,
  {Garc{\'{i}}a-Benito} R.,  {S{\'a}nchez} S.~F.,  {Husemann} B.,  {de Amorim}
  A.~L.,  {van de Ven} G.,  {Walcher} J.,  {Wisotzki} L.,  {Cortijo-Ferrero}
  C.,    {CALIFA Collaboration} 2013, \apjl, 764, L1

\bibitem[\protect\citeauthoryear{{Pilyugin}, {Grebel} \& {Kniazev}}{{Pilyugin}
  et~al.}{2014}]{P14}
{Pilyugin} L.~S.,  {Grebel} E.~K.,    {Kniazev} A.~Y.,  2014, \aj, 147, 131

\bibitem[\protect\citeauthoryear{{Pilyugin} \& {Thuan}}{{Pilyugin} \&
  {Thuan}}{2005}]{PT05}
{Pilyugin} L.~S.,  {Thuan} T.~X.,  2005, \apj, 631, 231

\bibitem[\protect\citeauthoryear{{Robles-Valdez}, {Carigi} \&
  {Peimbert}}{{Robles-Valdez} et~al.}{2013}]{r&v13}
{Robles-Valdez} F.,  {Carigi} L.,    {Peimbert} M.,  2013, \mnras, 429, 2351

\bibitem[\protect\citeauthoryear{{Rots}, {Bosma}, {van der Hulst},
  {Athanassoula} \& {Crane}}{{Rots} et~al.}{1990}]{rots90}
{Rots} A.~H.,  {Bosma} A.,  {van der Hulst} J.~M.,  {Athanassoula} E.,
  {Crane} P.~C.,  1990, \aj, 100, 387

\bibitem[\protect\citeauthoryear{{Salo} \& {Laurikainen}}{{Salo} \&
  {Laurikainen}}{2000}]{S&L00}
{Salo} H.,  {Laurikainen} E.,  2000, \mnras, 319, 377

\bibitem[\protect\citeauthoryear{{Scheepmaker}, {Lamers}, {Anders} \&
  {Larsen}}{{Scheepmaker} et~al.}{2009}]{scheepmaker09}
{Scheepmaker} R.~A.,  {Lamers} H.~J.~G.~L.~M.,  {Anders} P.,    {Larsen} S.~S.,
   2009, \aap, 494, 81

\bibitem[\protect\citeauthoryear{{Schruba}, {Leroy}, {Walter}, {Bigiel},
  {Brinks}, {de Blok}, {Dumas}, {Kramer}, {Rosolowsky}, {Sandstrom},
  {Schuster}, {Usero}, {Weiss} \& {Wiesemeyer}}{{Schruba}
  et~al.}{2011}]{schruba11}
{Schruba} A.,  {Leroy} A.~K.,  {Walter} F.,  {Bigiel} F.,  {Brinks} E.,  {de
  Blok} W.~J.~G.,  {Dumas} G.,  {Kramer} C.,  {Rosolowsky} E.,  {Sandstrom} K.,
   {Schuster} K.,  {Usero} A.,  {Weiss} A.,    {Wiesemeyer} H.,  2011, \aj,
  142, 37

\bibitem[\protect\citeauthoryear{{Schuster}, {Boucher}, {Brunswig}, {Carter},
  {Chenu}, {Foullieux}, {Greve}, {John}, {Lazareff}, {Navarro}, {Perrigouard},
  {Pollet}, {Sievers}, {Thum} \& {Wiesemeyer}}{{Schuster}
  et~al.}{2004}]{schuster04}
{Schuster} K.-F.,  {Boucher} C.,  {Brunswig} W.,  {Carter} M.,  {Chenu} J.-Y.,
  {Foullieux} B.,  {Greve} A.,  {John} D.,  {Lazareff} B.,  {Navarro} S.,
  {Perrigouard} A.,  {Pollet} J.-L.,  {Sievers} A.,  {Thum} C.,    {Wiesemeyer}
  H.,  2004, \aap, 423, 1171

\bibitem[\protect\citeauthoryear{{Schuster}, {Kramer}, {Hitschfeld},
  {Garcia-Burillo} \& {Mookerjea}}{{Schuster} et~al.}{2007}]{schuster07}
{Schuster} K.~F.,  {Kramer} C.,  {Hitschfeld} M.,  {Garcia-Burillo} S.,
  {Mookerjea} B.,  2007, \aap, 461, 143

\bibitem[\protect\citeauthoryear{{Theis} \& {Spinneker}}{{Theis} \&
  {Spinneker}}{2003}]{Theis03}
{Theis} C.,  {Spinneker} C.,  2003, \apss, 284, 495

\bibitem[\protect\citeauthoryear{{Tikhonov}, {Galazutdinova} \&
  {Tikhonov}}{{Tikhonov} et~al.}{2009}]{tikhonov09}
{Tikhonov} N.~A.,  {Galazutdinova} O.~A.,    {Tikhonov} E.~N.,  2009,
  \AstronomyLetters, 35, 599

\bibitem[\protect\citeauthoryear{{Tinsley}}{{Tinsley}}{1980}]{Tinsley80}
{Tinsley} B.~M.,  1980, \fcp, 5, 287

\bibitem[\protect\citeauthoryear{{Toomre} \& {Toomre}}{{Toomre} \&
  {Toomre}}{1972}]{Toomre2}
{Toomre} A.,  {Toomre} J.,  1972, \apj, 178, 623

\bibitem[\protect\citeauthoryear{{Tully}}{{Tully}}{1974}]{tully74}
{Tully} R.~B.,  1974, \apjs, 27, 437

\bibitem[\protect\citeauthoryear{{Walter}, {Brinks}, {de Blok}, {Bigiel},
  {Kennicutt} Jr., {Thornley} \& {Leroy}}{{Walter} et~al.}{2008}]{walter08}
{Walter} F.,  {Brinks} E.,  {de Blok} W.~J.~G.,  {Bigiel} F.,  {Kennicutt} Jr.
  R.~C.,  {Thornley} M.~D.,    {Leroy} A.,  2008, \aj, 136, 2563

\bibitem[\protect\citeauthoryear{{Yin}, {Hou}, {Prantzos}, {Boissier}, {Chang},
  {Shen} \& {Zhang}}{{Yin} et~al.}{2009}]{yin09}
{Yin} J.,  {Hou} J.~L.,  {Prantzos} N.,  {Boissier} S.,  {Chang} R.~X.,  {Shen}
  S.~Y.,    {Zhang} B.,  2009, \aap, 505, 497

\bibitem[\protect\citeauthoryear{{Zwicky}}{{Zwicky}}{1955}]{Zwicky55}
{Zwicky} F.,  1955, \pasp, 67, 232

\end{thebibliography}
\bibliographystyle{mn2e}

\label{lastpage}
\end{document}